Comparison of 2D simulation models to estimate the critical current of a coated superconducting coil

Yingzhen Liu, Jing Ou, Francesco Grilli, Fabian Schreiner, Victor Manuel Rodriguez Zermeno, Min Zhang and Mathias Noe

Abstract
Superconductors have been being applied to a variety of large-scale power applications, including magnets, electric machines, and fault current limiters, because they can enable a compact, lightweight and high efficiency design. In applications such those mentioned above, superconducting coils are always a key component. For example, in a superconducting electric machine, the superconducting coils are used to generate the main flux density in the air gap, which is significantly important for the energy conversion. It is the performance of the superconducting coils that plays an essential role in determining the performance of the device. However, the performance of a superconducting coil is limited by its critical current, which is determined by temperature and the magnitude and orientation of the magnetic field inside the superconductors. Hence, in-depth investigations to estimate the critical current of the superconducting coils are necessary before manufacturing. Available transient simulation models to estimate the critical current are through the *H*- and *T-A* formulations of Maxwell's equations. Both methods consider the same current ramp-up process occurring in experiments. Besides these transient models, static simulations can also be used: a modified load-line method and the so-called *P*-model, which is based on the asymptotic limit of Faraday's equation when time approaches infinity. To find the best way to calculate the critical current, the four methods are used to estimate the critical current of a double pancake superconducting coils and results are compared with experiments. As a conclusion, *T-A* formulation, *P*-model, and the modified load-line methods are recommended for estimating the critical current of the superconducting coils.

I. Introduction

The unique properties possessed by (RE)BaCuO (RE=rare earth) coated conductors, high current carrying ability, relatively high operational temperature, and good in-field performance, have broadened the approaches to design magnets and electric machines. Many improvements have been made in (RE)BaCuO conductors, resulting in better conductor quality and longer available length. Currently, the production capacity of (RE)BaCuO has reached 1000 km/year and its critical current at 77 K, self-field is around 100-200 A with 4 mm width and 200-600 A with 12 mm width; all over the world there are more than 12 companies which produce (RE)BaCuO superconductors [1],[2]. The price of the superconductors is around 100-1000 $/kA-m (using critical current at 77 K, self-field). The range of the price depends on manufacturers and volume of customers' order.

By applying (RE)BaCuO conductors to large scale power devices, a compact and light weight design with a higher current or higher magnetic field capacity than conventional materials can be achieved. For example, a 10 MW superconducting direct-drive wind generator can have a double torque-to-mass ratio and a quadruple torque-to-volume ratio, compared to the permanent magnet generators at the same rating due to higher magnetic field is produced [3]. Because of the layered structure of the coated conductors, stacks of pancake coils are employed in the magnets and superconducting motors and generators. The electromagnetic behaviours of the pancake coils are crucial in determining the performance of superconducting device. Accurately evaluating the maximum current that the coil can safely carry is of high importance.

Numerically, the finite element method (FEM) based on *H*-formulation has been used for modelling the critical current of pancake coils. M. Zhang *et al.* [4] studied the critical current of a 40-turn circular YBaCuO pancake coil by setting up a 2D axisymmetric model in *Comsol* based on *H*-formulation. M. D. Ainslie *et al.* [5] used the same method to investigate the in-field critical current of circular pancakes with 20 and 50 turns. With this method, the real thickness of the superconducting layer is modelled, even though there is high aspect ratio of the tape dimension. However, when the number of turns of the pancake coils increases to hundreds, this FEM approach based on the *H*-formulation will suffer from long computation time [6]. For instance, in the case of a 10 MW class wind generator, the number of turns per double pancake racetrack superconducting coil is designed to be up to 2000 [7]-[10]. Reducing the computation time is therefore necessary.

To speed-up the computation time of dynamic problems considering large number of turns, H. Zhang *et al.* [11] and F. Liang *et al.* [12] proposed an efficient transient FEM based on the *T-A* formulation. In [11] the ac losses of a 5-turn racetrack pancake coil were calculated in 3D and in [12] the ac losses of a 78-turn double pancake racetrack coil and 2000-turn pancake coils in 2D were computed. With the *T-A* formulation, the thickness of the superconductors is ignored: the superconductor layers are approximated to sheets in 3D and lines in 2D to tackle the high aspect ratio problem. In this way, a pancake coil with large number of turns can be modeled in a short time. However, there is a lack of experimental validation for the large number of turns' model. Besides, this method can be only applied to coils made by coated superconductors.

V. M. R. Zermeno *et al.* [6] proposed a steady-state model to evaluate the critical current of cables and coils for all types of superconducting wires and tapes. The model, known as *P*-model after the name of an auxiliary variable used for the calculations, assumes that the magnetic field and current density will reach a steady state when time goes to infinity with dc excitation. Correspondingly, the auxiliary variable *P* and electric field are uniform over the conductor's cross section. As there is no ramping process from zero initial conditions in the calculation, the computation is very fast. Readers should note that in the *P*-model the true dynamics behind the voltage drop at current density values that are far from the critical value are not considered. C.R. Vargas-Llanos *et al.* [13] used it for estimating the critical current of multi-filamentary $MgB_2$ tapes and D. Liu *et al.* [14] used this model to estimate the critical current of Bi-2223 cables and coils.

In the engineering field, the well-known load-line method is used to estimate the critical current when the superconductor is applied to cables [15], transformers [16], magnets [17], motors and generators [18], and so on. Briefly, the load-line method calculates the magnetic field of a superconducting coil as a function of the current and compares it to the single tape critical current as a function of applied field. The crossing point of the two lines is the critical current of the coil. In this method, the important assumption is that the current density inside the superconductor is uniform, which is not the case in reality.

The objective of this paper is to investigate the critical current of a superconducting coil with large number of turns by using the above mentioned models, to compare the available models for this coil in terms of computation efficiency and accuracy with respect to experiment results, and to suggest the most efficient way to estimate critical current. The paper is organized as follows. In section 2, the 2D models are briefly introduced. In section 3, the anisotropic field dependence of the critical current of the tape is described and the measurements of the critical current of a 244-turn double pancake racetrack coil are summarized, followed by the comparison and discussion of the models in section 4. Finally, the conclusion is drawn in section 5.

II. 2D models

In order to simplify the geometry, the racetrack pancake coil is modelled in 2D infinitely long model [12],[19] and the end effect of the two half circles are neglected as illustrated in Fig. 1. This simplification is made because the length of the coil is much longer than the diameter of the circle. In this coil, each turn has the same supplied current. Furthermore, based on the symmetry of the coil, only one leg of the coil is simulated.

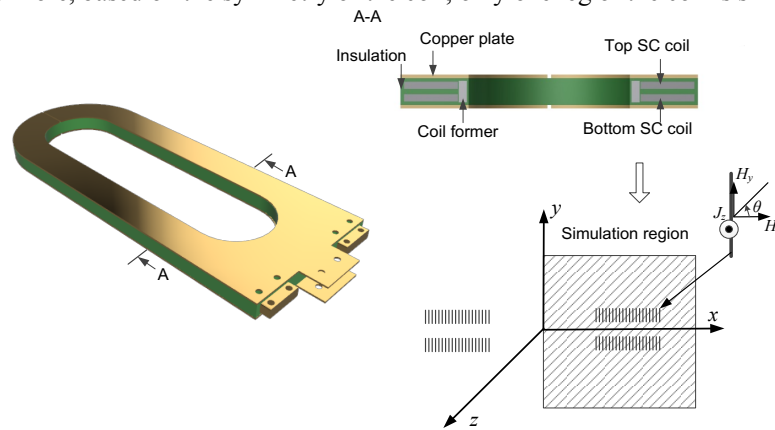

Fig. 1 Sketch of the modeled superconducting (SC) double pancake racetrack coil.

A. *H*-formulation

The governing equation of the *H*-formulation is derived from Faraday's law and Ampere's law and it directly solves the magnetic field components $H=[H_x,H_y]$ in 2D geometry in *x-y* plane. With non-magnetic subdomains, such as the superconductor and air, the corresponding partial differential equations are described by [20]-[23],

$$\frac{\partial E_z}{\partial x} = \mu_0 \frac{\partial H_y}{\partial t} \quad (1)$$

$$\frac{\partial E_z}{\partial y} = -\mu_0 \frac{\partial H_x}{\partial t} \quad (2)$$

$$J_z = \frac{\partial H_x}{\partial y} - \frac{\partial H_y}{\partial x} \quad (3)$$

where $H_x$, $H_y$ represent the component of magnetic field in *x* and *y* directions. The current density and electric field have only *z*-component, denoted by $J_z$ and $E_z$, $\mu_0$ is the permeability of vacuum, and *t* is time.

To describe the constitutive relation of materials, the *E-J* relation is expressed as follow,

$$E_z = \rho J_z \quad (4)$$

where *ρ* represent the resistivity of materials. Vacuum is described by insulator with zero conductivity, and the resistivity of the superconducting material, $\rho_{sc}$, is given by the *E-J* power law [24]:

$$\rho_{sc} = \frac{E_0}{J_c(B_x, B_y)} \left( \frac{J_z}{J_c(B_x, B_y)} \right)^{n-1} \quad (5),$$

where $E_0$ is the critical electric field, which is 1 $\mu$V/cm, and $n$ is a parameter indicating the sharpness of the transition from superconductor to the normal state, $J_c$ is the critical current density, $B_x$ and $B_y$ are the two components of the magnetic field density in $x$ and $y$ directions.

The critical current is determined based on the choice of critical current $I_c$ criteria, which can vary accordingly to the application. The two different criteria are [6]:

(i) MAX criterion, $I_c$ is the current at which the voltage drop per unit length has reached the critical electrical field in at least one turn.

(ii) SUM criterion, $I_c$ is the current at which the sum of the voltage drop in each turn, divided by the coil's length, reaches the critical electrical field.

In this paper, only the SUM criterion is applied to estimate the critical current of the coil to match the experiment conditions. To implement this in the program, the average electrical field in each superconductor is multiplied by each turn's length, and then the sum of all the products is divided by the total coil's length. When this value reaches the critical electric field, the critical current is determined.

B. *T-A* formulation

In a 2D geometry in $x$-$y$ plane, the *T-A* formulation uses two state variables: the current vector potential $T$, which only has $x$ component, $T_x$, and the magnetic vector potential, which only has $z$ component, $A_z$, when the thickness of the superconducting layer in the $x$ direction is ignored. The current vector potential is calculated in the superconductor, while the magnetic potential is calculated in the whole region. The governing equations in the two regions are described by [11],[12]:

$$\frac{\partial T_x}{\partial y} = J_z \quad (6)$$

$$\frac{\partial E_z}{\partial y} = \frac{\partial B_x}{\partial t} \quad (7)$$

$$\frac{\partial^2 A_z}{\partial x^2} + \frac{\partial^2 A_z}{\partial y^2} = \mu_0 J_z \quad (8)$$

$$B_x = -\frac{\partial A_z}{\partial y} \quad (9)$$

The current density distribution $J_z$ in the superconductor and the magnetic flux density $B_x$ are exchanged by the two formulations, where $B_x$ represents the component of magnetic field density in the $x$ direction. Then the same process to determine the critical current of the coil is used as the *H*-formulation.

C. *P*-model

In the P-model, the parameter $t$ disappears as the model assumes that the magnetic flux density and current density have reached the steady state with dc excitations when $t$ goes to infinity. This model introduces a parameter $P$ to rewrite the *E-J* power law, Eq. (5), into [6],[14].:

$$J_z = J_c(B_x, B_y) P \quad (10)$$

$$P = \left( \frac{E_z}{E_0} \right)^{\frac{1}{n}} \quad (11)$$

The introduction of parameter $P$ avoids the direct solution of the nonlinear *E-J* relationship for the superconductor. In the cross section of each conductor, $P$ is uniformly distributed and is given by

$$P = \frac{I_z}{\int_S J_c(B_x, B_y) dxdy} \quad (12)$$

where $I_z$ is the transport current of the superconductor, $S$ is the cross section of the superconductor, and $J_c(B_x,B_y)$ represents the critical current density distribution inside the superconductor. To calculate the magnetic field density, the magnetic vector potential is used and Eq. (8) can be rewritten as Eq. (13). The $x$ and $y$ component of the magnetic flux density are calculated by Eq. (9) and Eq. (14).

$$\frac{\partial^2 A_z}{\partial x^2} + \frac{\partial^2 A_z}{\partial y^2} = \mu_0 J_c(B_x, B_y) P \quad (13)$$

$$B_x = \frac{\partial A_z}{\partial y} \quad (14)$$

The termination of the program is made based on the SUM criterion, the same as *H*- and *T-A* formulation.

D. Modified load-line method

The load-line method assumes uniform current density distribution inside the superconductor layer. According to Eq. (8), (9) and (14), at a certain current, the magnetic field inside one conductor is calculated. By using the $I_c(B)$ of the short tape sample, the critical current of the conductor is calculated at this given current. The simulation is repeated for increasing values of the applied current. When the calculated critical current is the same as the given current, the maximum current the conductor can carry is determined. In this paper, we modified the load-line method as follows: for each given current, the magnetic field distribution inside the superconductors is calculated as aforementioned; according to the $J_c(B_x,B_y)$ of the short tape sample and the calculated flux density distribution, the critical current density distribution of the conductor is calculated at this given current; instead of finding the crossing point of the given currents and critical currents, the average electric field of each turn is calculated according to Eq. (4) and (5); afterwards, the same process to determine the critical current as in the *H*-formulation is employed. Because of the assumption of uniform current density distribution to calculate the magnetic flux density, this method is named as modified load-line method.

III. Measurement of the tape and the double pancake racetrack coil

A. Anisotropy of the tape

The anisotropy of the tape's critical current under an applied field will greatly influence the critical current. Hence, we first measured the lift factor of a short sample as a function of the orientation and magnitude of the applied field. The lift factor is defined as the ratio of the critical current at a certain orientation and amplitude of the magnetic field and temperature to the critical current at 77 K with self-field. The tape we used to wind the double pancake racetrack coil is 4 mm in width and the minimal critical current of the tape along overall length is 100 A according to the manufacturer's specification. The measured angular and field dependence of the superconducting tape in liquid nitrogen is shown in Fig. 2. The $\theta=0°$ is defined as perpendicular field to the tape's wider surface. At position 90°, the Lorenz force pushes the superconductor layer towards the sample surface (opposite direction to the substrate). From Fig. 2 it can be seen that the highest critical current is obtained at around the angle of 65° and 245° instead of parallel field (90°, 270°) to the wider surface. This is because the superconducting crystals are grown with a 25° shift to the perpendicular field [25]. The curves in Fig. 2 are almost 180°-symmetric and the slightly difference of the maxima at 65° and 245° is explained by the stronger surface barrier on the superconductor-air interface than on the superconductor-substrate surface [26],[27].

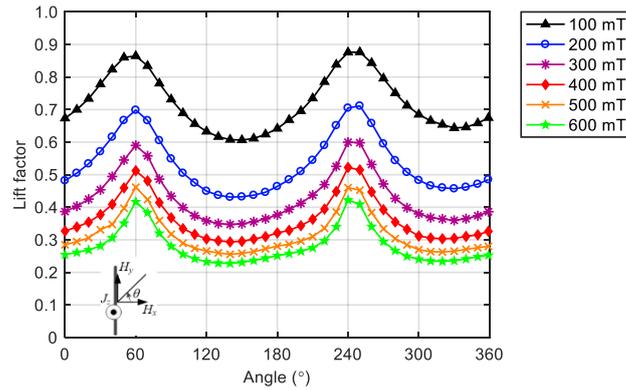

Fig. 2 Lift factor of superconducting tape TPL2300c as a function of orientation and amplitude of an applied field in liquid nitrogen at 77 K

In order to accurately estimate the critical current of the coils, it is necessary to know the *n*-value of the tape in the power-law equation (5). To derive the *n*-values, the measured current-voltage data points of the tape between 0.35 $\mu$V/cm and 1 $\mu$V/cm are employed for a given applied field. Based on Eq. (5), the derived *n*-value is shown in Fig. 3. From Fig. 3 it can be seen that, in general, the higher the magnetic flux density, the lower the *n*-value. The variation of the *n*-value is between 15.1 and 36.4 for magnetic flux density in the range of 100-600 mT. In addition, it is very difficult to extract a sensible $n(B_x,B_y)$ relation from these data in Fig. 3. For simplification, when the critical current of the double pancake racetrack coil is simulated, a constant *n*-value is used and the influence of different *n*-values on determining the critical current of the coil is investigated in section IV.

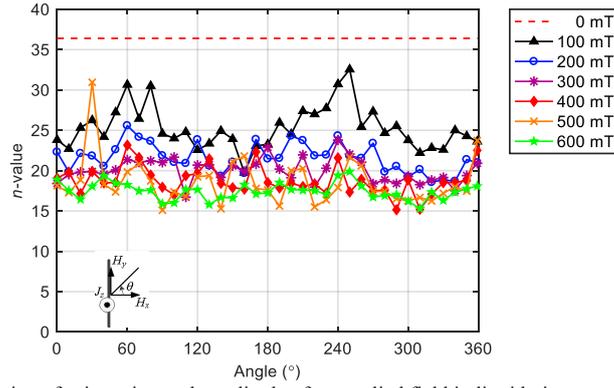

Fig. 3 Derived *n*-values as a function of orientation and amplitude of an applied field in liquid nitrogen at 77 K

B. Critical current measurement of the double-pancake racetrack coil

The structure of the double pancake racetrack coil used in this paper is shown in Fig. 1 and a photo of the coil is shown in Fig. 4. The specification of the coil is summarized in Table I. The double pancake coil consists of 244 turns of GdBaCuO tapes. The top and bottom coil are insulated by a 3 mm-thick insulator. The turn-to-turn insulation is co-wound with the GdBaCuO tapes on a coil former, which is used for mechanical stabilization. The layer-to-layer connection is achieved by soldering a piece of 12 mm GdBaCuO tape. Afterwards, the coil is wrapped by resin. Two copper plates are placed at the top and bottom to enhance the thermal conductivity in conduction cooling.

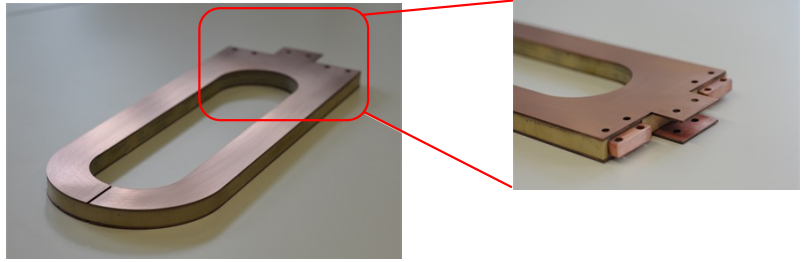

Fig. 4 Photo of the SC double pancake coil.

Table I. Specification of the superconducting double pancake coil

| Parameter | Value |
| --- | --- |
| Tape | GdBaCuO |
| Tape thickness | 0.22 mm |
| Tape width | 4 mm |
| Copper stabilizer | 100 um |
| Min. tape $I_c$ (77 K, sf.) | 100 A |
| Number of turns | 2×122 |
| Thickness of copper plate | 2 mm |
| Height of the coil | 19 mm |
| Inner/outer radius of the coil | 45/83 mm |
| Total length of tapes | 218 m |

To measure the critical current of the coil, the four point measurement technique is employed. The current is slowly ramped up at the terminals of the coil, and the voltage between the terminals is recorded. When the voltage reached the given value of the electrical field, the corresponding current is the so-called critical current. In general, two critical electrical fields are used for determining the critical current, which are 0.1 $\mu$V/cm or 1 $\mu$V/cm. In this paper, both criteria are considered. In order to observe the voltage of the double pancake coil, we soldered three voltage taps: two in the terminals and one in the middle of the whole coil. The tested *V-I* curves of the superconducting double pancake coil with a criterion of 0.1 $\mu$V/cm in liquid nitrogen at 77 K, and by the cryogen-free cooling method down to 77 K are illustrated in Fig. 5, respectively. To achieve the cryogen-free cooling method, the cold head of the cryocooler is attached to the two copper plates.

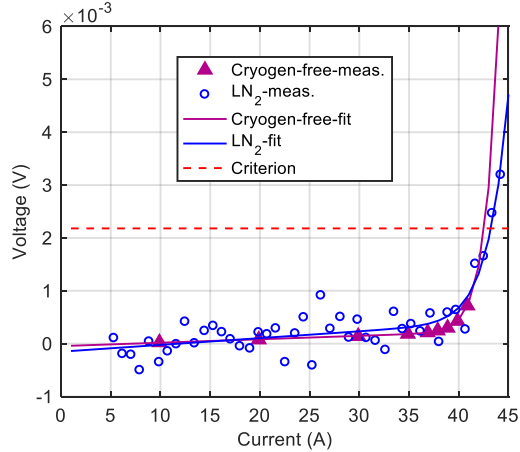

Fig. 5 Measured *V-I* curves of the double pancake superconducting coil in liquid nitrogen bath, and by cryogen-free cooling method down to 77 K, respectively.(note: criterion here is 0.1 $\mu$V/cm)

The critical current of the double pancake coil is 43.2 A in liquid nitrogen, and 42.6 A by cryogen-free cooling method with the critical electric field 0.1 $\mu$V/cm. The derived *n*-value in the power law in liquid nitrogen is 22.1, while it is 33.3 by cryogen-free cooling. The difference of the *n*-values is caused by the different cooling methods. In the liquid nitrogen bath, there is a high heat transfer due to nucleate boiling and an almost constant surface temperature of the coil is obtained. Differently, through the cryocooler, a thermal link, in this case the two copper plates, is necessary to distribute the cooling power over the whole coil. The thermal diffusion in the thermal link and the contact thermal resistance between the tapes and copper plates will limit the cooling for transient state. When we ramped up the current in the coil close to the critical current, due to the small ohmic losses, the transient temperature is slightly higher by the cryogen-free cooling than in liquid nitrogen bath, which contributes to a higher electrical field at this certain current. As shown in Fig. 5, for the same input current, the voltage of the coil by cryogen-free cooling method is higher when the current is close to the critical current. According to Eq. (5), a higher *n*-value is derived.

With the 1 $\mu$V/cm criterion, the critical current in liquid nitrogen cooling is 48.7 A. For the case of cryogen-free cooling, due to the weak thermal coupling between the copper plates and the superconducting tapes, the coil has already reached instability (thermal runaway quench) before the criterion. Hence, we use the fitting curve value to estimate the critical current with criterion 1 $\mu$V/cm, which is 45.8 A (this is not visible in Fig. 5).

The tested *V-I* curves of the whole double-pancake coil, the bottom coil and the top coil, as shown in Fig. 1 are illustrated in Fig. 6. From Fig. 6 it can be seen that when the top coil reaches its criterion, the bottom coil is still in the superconducting state, that is, the top coil determines the critical current. This will discussed in Section IV-A.

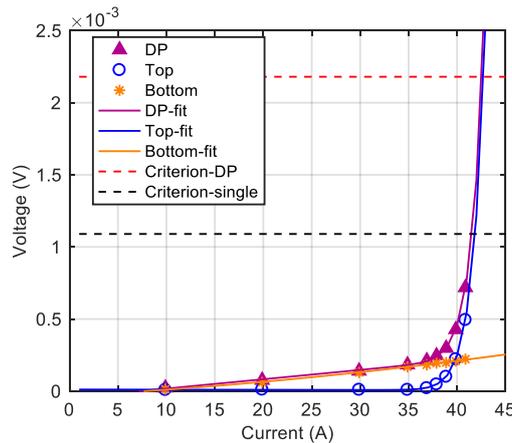

Fig. 6 Measured *V-I* curves of the double pancake coil, top coil and bottom coil by cryogen-free cooling method (note: DP is abbreviation of double pancake)

IV. Comparison and discussion of the simulation models

　A.　Current density and magnetic field distribution

To simulate the above mentioned double pancake coil in *Comsol*, the local $J_c(B_x,B_y)$ dependence of the tape is needed as input of the calculation. Local $J_c(B_x,B_y)$ means the dependence of the critical current density on the local magnetic flux density, resulting from both the applied external field and self-field. This can be extracted from the experiment data shown in Fig. 2 by means of the parameter-free model presented in [28]. There is a huge computation time difference for the four models. With *Comsol* 5.3a the *H*-formulation takes about 10 hours, while the *T-A* formulation needs 8 minutes, the *P*-model 5 minutes and the modified load-line method 3 minutes (the processor of the computer is Intel(R) Core(TM) i7-4930K CPU@3.40 GHZ, and the installed RAM is 16.0 GB). Regarding the speed, the latter three models are favorable.

The main differences of the four models, besides the stationary or transient studies, are the following: in the *H*-formulation and *P*-model, the current density distributes along the tape's width and thickness; in the *T-A* formulation the current density distributes along the tape's width, since the tape's thickness is neglected; the modified load-line method assumes uniform current density distribution within the tape's cross section. Correspondingly, the local magnetic field and critical currents vary for the four models. The magnetic field distributions and the average $J/J_c$ of each turn of the double pancake coil calculated by the four models are compared in Fig. 7 and Fig. 8 with applied current of 51 A and a fixed *n*-value of 30. Even though 51 A is larger than the measured critical current of the coil, it is still below but close to the critical current in the simulation with the criterion of 1 $\mu$V/cm. From Fig. 7 it can be seen that, on the cale o the whole coil, the magnetic field flux density distributions are similar for the four models. From Fig. 8 it can be seen that the critical currents are also similar for all four models. The largest deviation of the critical current of the *T-A* formulation is 2.6%, 1.8% of the *P*-model and 0.8% of the modified load-line model compared to that of the *H*-formulation.

Form Fig. 8 it can be seen that in all models the innermost turn of the top coil has the lowest critical current and the top coil reaches its critical current earlier than the bottom coil, which is consistent with the measurement in Fig. 6. In addition, with the 1 $\mu$V/cm criterion, the innermost turns of the top coil have already been over their critical current. It is the anisotropy of the tape, as shown in Fig. 2, that causes the top coil to have a lower critical current. From Fig. 7 it can be noted that the magnetic flux densities at the top and bottom coils have same the amplitude, but opposite field angles, when zero angle is defined as the perpendicular field to the tape's wider surface. The fact that the top coil has lower critical current is consistent with the conclusion in [26]. However, in [26] the two critical currents for the top and bottom coils are only slightly different. The relatively large difference in our measurement can be explained by the 25 degree shift.

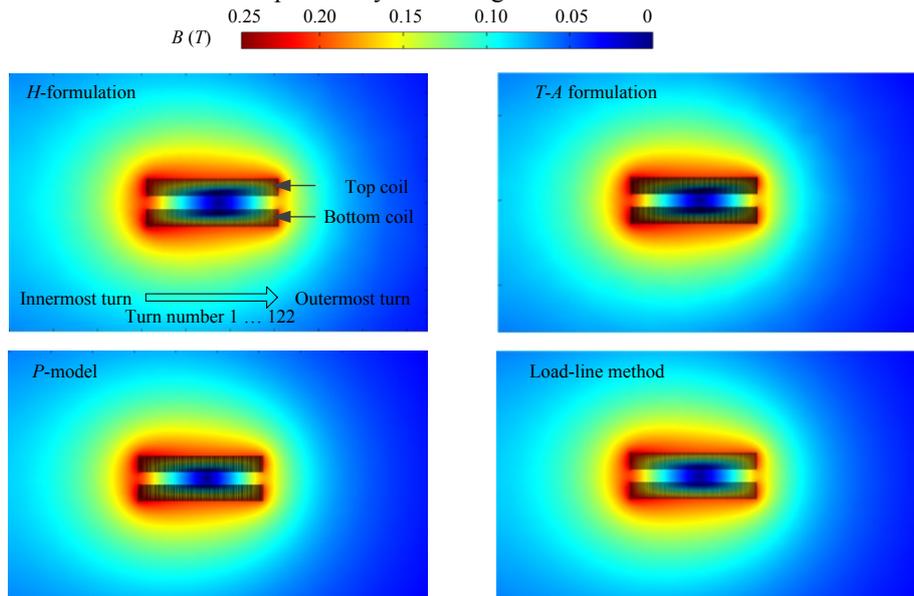

Fig. 7 Calculated magnetic field distributions of the double-pancake coil at the current of 51 A

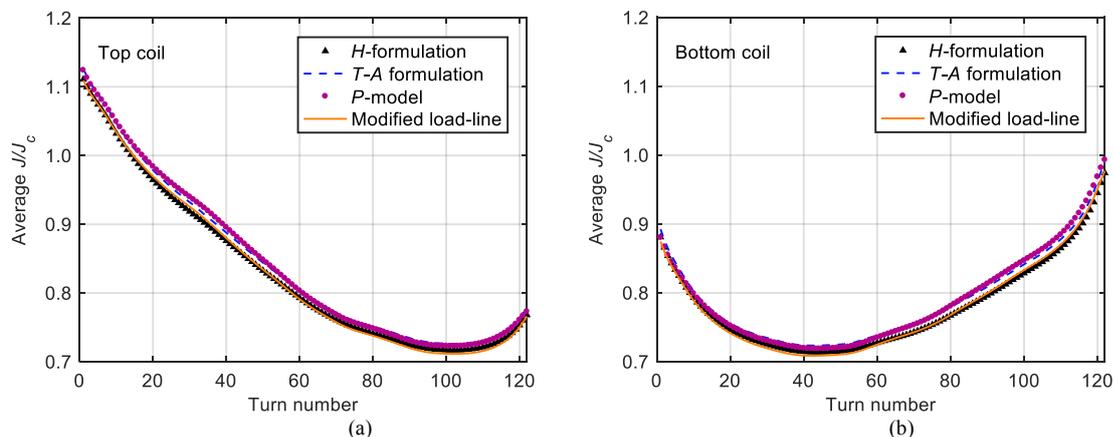

(a)          (b)

Fig. 8 Calculated average $J/J_c$ of each turn in the double pancake coil at the current of 51 A, (a) the top coil, and (b) the bottom coil

*B. n-value influence*

As illustrated by Eq. (5), the *n*-value is a key parameter to simulate the superconductors. Different *n*-values will affect the critical current estimation. From Fig. 3 it can be seen that the *n*-value changes with the amplitude and the orientation of the magnetic field. In this section, the *n*-value's effects are investigated and compared to the measurement results. Fig. 9 shows the influence of *n*-values when the critical current criterion is 1 $\mu$V/cm and 0.1 $\mu$V/cm. The critical current increases with the *n*-values when the criterion is 0.1 $\mu$V/cm, whereas decreases with the *n*-values with criterion 1 $\mu$V/cm. The exception of *P*-model with criterion of 0.1 $\mu$V/cm can be explained by the assumption of the model. There is no critical electric field in the governing Eq. (13): the criterion of the coil is intrinsically decided by the criterion used for the characterization of the sample, while the criterion of other models is not influenced by the criterion chosen by the tape characterization. Therefore, the trends of the influence of *n*-values with both criteria are the same for the *P*-model. The estimated critical currents from *T-A* formulation are the same as those from *H*-formulation, but the computation is much faster. Fig 9 also illustrates that the larger the *n*-value, the smaller the difference of the critical currents between the criteria of 0.1 and 1 $\mu$V/cm, which is reasonable due to a sharper transition of the superconductor. The critical current calculated by the modified load-line method is the smallest. No matter which model we choose, the calculated value is always overestimated compared to the measurement and reasons are discussed in the following section. When the *n*-value, as illustrated in Fig. 3, ranges from 15-40, it can be seen from Fig. 9 that the critical current of the double pancake superconducting coil calculated by the models is within 19% of the measured value. The maximum deviations compared to the measurement are summarized in Table II and Table III for the *n*-value ranging from 15-40. The modified load-line method seems to have a higher accuracy in this case as it has lower critical current estimation than the others.

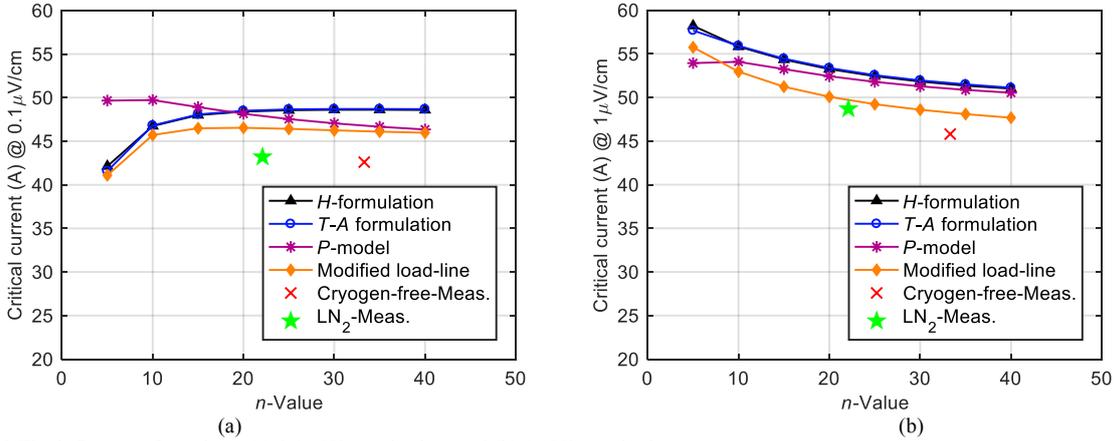

Fig. 9 The influence of n-values, (a) 0.1 $\mu$V/cm criterion, and (b) 1 $\mu$V/cm criterion.

Table II. Maximum deviation of calculated critical currents compared to cryogen-free cooling measurement

| Criterion | *H*-formulation | *T-A* formulation | *P*-model | Modified load-line |
|---|---|---|---|---|
| 0.1 $\mu$V/cm | 14.1% | 14.3% | 14.8% | 9.1% |
| 1 $\mu$V/cm | 18.7% | 18.9% | 16.3% | 11.2% |

Table III. Maximum deviation of calculated critical currents compared to liquid nitrogen measurement

| Criterion | *H*-formulation | *T-A* formulation | *P*-model | Modified load-line |
|---|---|---|---|---|
| 0.1 $\mu$V/cm | 12.5% | 12.7% | 13.2% | 7.7% |
| 1 $\mu$V/cm | 11.2% | 11.8% | 9.3% | 5.2% |

*C. Summary of the simulation models*

As presented above, the models agree with experiments within 5%-19%, and they all overestimate the critical current. The fact that the modified load-line method has 'better' agreement should be taken cautiously, because there are other factors that can lead to this. One possible reason is that in this simulation a fixed *n*-value is used for the superconductors, while in reality it changes with the amplitude and the orientation of the magnetic field, $n(B_x,B_y)$. A more important factor, however, is the tape uniformity. Currently, the uniformity of the properties of commercial available superconductors along length depends on manufacturing process [29] and it is challenging to guarantee the same critical current for the total length needed for a superconducting coil with large number of turns. For instance, a Tapestar measurement of a short sample used in the coil at 77 K, self-field is illustrated in Fig. 10. Along the 0.7 m length, the critical current varies 45% compared with the minimal value. In [30] it was shown that a long-length superconducting tape with a deviation of 10% can contribute to a 38% lower critical current of a Roebel cable than that of a cable made of strands with identical transport capacity. Even though in

[30] Roebel cable is discussed, not the superconducting coil, that work shows that the length uniformity of the tape has a big influence on the critical current of superconducting devices. Another important factor is that the length uniformity of the in-field angular dependence of the superconductor. In [31] it has been pointed out that the length uniformity of the angular dependence of the critical current is between 0.3% and 6.7% for SuperPower tape with artificial pinning centers up to 600 mT. Moreover, the manufacturing process of the superconducting coil, for instance, winding and impregnation, influences the measured critical current. Even though it is hard to evaluate the influence, the manufacturing process can lead to certain degradation of the tapes and the insulation and resin do matter in the critical current measurement. Last not the least, the cooling of a single tape used for characterization of $I_c$ is much more efficient than that of a coil, even in a liquid nitrogen bath. All the reasons contribute to the inaccuracy of critical current estimation by numerical models.

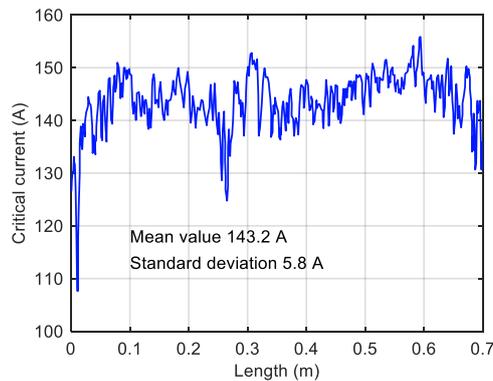

Fig. 10 Length uniformity of critical current of the short sample at 77 K from Tapestar measurement.

Compared to the measured critical currents, the modified load-line gives result closer to experiment, not because it explains what happens in the superconductor, but because uniform current density assumption gives lower $I_c$ estimation. If ideal tapes are used, the other three models should give better results. In conclusion, the *T-A* formulation, *P*-model and modified load-line method are recommended to estimate the critical current of a coated superconductor coil in terms of computation efficiency, compared to *H*-formulation. The *T-A* formulation is effective to estimate critical current and losses for the coated conductors. The *P*-model and modified load line methods can be used with all type of superconductors. Especially for coils in magnets or electric machines, the modified load-line method is easy to implement in operating conditions together with all components including ferromagnetic materials. However, the applicability of the modified load-line method to more complex coils (e.g. coils made of Roebel cable) has to be verified. In this case, the *P*-model is favorable, but the criterion to decide the critical current of the superconducting coil is inherently determined by the criterion used for the characterization of the short sample.

Conclusion

This work is dedicated to comprehensively compare available numerical models, *H*-formulation, *T-A* formulation, *P*-model, and modified load-line method, to estimate the critical current of a double pancake racetrack coil with more than 200 turns. By implementing the characterization of the short sample into the simulation models, the critical currents are calculated. In order to evaluate the calculated critical currents, a double pancake coil was manufactured and tested at 77 K both in liquid nitrogen bath and by cryogen-free cooling method. The critical current by the cryogen-free cooling is 1.4% smaller with criterion of 0.1 $\mu$V/cm and 6.0% smaller with 1 $\mu$V/cm compared to that in liquid nitrogen bath. In particular, the *n*-value at a criterion of 0.1 $\mu$V/cm changes from 33.3 to 22.1, due to different cooling conditions. It is suggested to operate the coil at a relatively smaller current by the cryogen-free cooling than by the liquid nitrogen bath cooling.

Compared to experimental results, the four models overestimated the critical current of the coil, within 19% when the *n*-value ranges from 15 to 40. This deviation is very likely caused by the varying *n*-values, length uniformity of critical current in field and angular dependence, manufacturing process as well as the different cooling efficiency between the short sample and the whole superconducting coil. The attributes of the four methods are listed in Table IV. In summary, it is fair to recommend the *T-A* formulation, the *P*-model and the modified load-line method to estimate the critical current of a coated superconducting coil with large number of turns.

Furthermore, consistent with the measurement, all four models show that the top coil of the double pancake superconducting coil reaches its critical current earlier than the bottom coil due to the anisotropy of the critical current in field. Last not the least, with the 1 $\mu$V/cm criterion, the innermost turns of the top coil are already over critical current. Therefore, it is better to determine the critical current by the 0.1 $\mu$V/cm criterion to ensure safe operation of the double pancake coil with large number of turns.

Table IV. Attributes of the four models

|  | H-formulation | T-A formulation | P-model | Modified load-line method |
|---|---|---|---|---|
| Possible applicability | All types of superconducting wires/tapes | Coated superconductors | All types of superconducting wires/tapes | All types of superconducting wires/tapes |
|  | Cables and coils | Cables and coils | Cables and coils | Coils with large number of turns |
| Speed | Several hours | Few minutes | Few minutes | Few minutes |

Acknowledgement

We acknowledge the support of Peter van Hasselt and Tabea Arndt, Siemens AG, Corporate Technology Erlangen, for hosting in the lab, supplying the equipment and for specific advice in performing the measurements on the coil. We would like to thank Markus Bauer, THEVA Dünnschichttechnik GmbH, for supports of manufacturing the tape and coil.

This work is partly supported by German Research Foundation under Grant NO 935/1-1, and the National Natural Science Foundation of China under Grant 51761135120.